\input{psfig.sty}  
  \documentclass{aa}  
\usepackage{txfonts}
\usepackage{psfig}
 \usepackage{rotating}
\usepackage{subfigure}
\usepackage{aas_macros}
\usepackage{amssymb}
\usepackage{natbib}
\bibpunct{(}{)}{;}{a}{,}{,}
\begin{document}
\title{Circumstellar Disks in the Outer Galaxy: the Star-Forming Region NGC 1893\thanks{Tables \ref{table_c} and \ref{member_cat} are only available in electronic form
at the CDS via anonymous ftp to cdsarc.u-strasbg.fr (130.79.128.5)
or via http://cdsweb.u-strasbg.fr/cgi-bin/qcat?J/A+A/}}
\author{M. Caramazza\inst{1,2}
          \and G.
	  Micela\inst{2} \and L. Prisinzano \inst{2}  \and L. Rebull \inst{3} \and S. Sciortino \inst{2} \and J. R. Stauffer
	 \inst{3}}
	 \institute{Dipartimento di Scienze Fisiche ed Astronomiche, Universit\`a di Palermo , Via Archirafi 36, 90123, Palermo, Italy \and INAF Osservatorio Astronomico di Palermo, Piazza del Parlamento 1, 90134 Palermo, Italy \and \textit{Spitzer} Science Center, Caltech 314-6, Pasadena, CA 91125\\ 
	  email:mcarama@astropa.unipa.it}	  
\date{Received 2008 May 05; accepted 2008 May 30.}

 \abstract
{It is still debated whether star formation process depends on environment. In particular it is yet unclear whether star formation in the outer Galaxy, where the environmental conditions are, theoretically, less conducive, occurs in the same way as in the inner Galaxy.}
{We investigate the population of NGC1893, a young cluster $(\sim 3-4$ Myr) in the outer part of the Galaxy (R$_G$ $\geq$ 11 Kpc), to explore the effects of environmental conditions on star forming regions.}
{We present infrared observations acquired using the IRAC camera onboard the \textit{Spitzer} Space Telescope and analyze the color-color diagrams to establish the membership of stars with excesses. We also merge this information with that obtained from \textit{Chandra} ACIS-I observations, to identify the Class III population.}
{We find that the cluster is very rich, with 242 PMS Classical T-Tauri stars and 7 Class 0/I stars. We identify 110 Class III candidate cluster members in the ACIS-I field of view. We estimate a disk fraction for NGC1893 of about $67 \%$, similar to fractions calculated for nearby star forming regions of the same age.}
{Although environmental conditions are unfavorable, star formation can clearly be very successful in the outer Galaxy, allowing creation of a very rich cluster like NGC1893.}
\titlerunning{Circumstellar Disks in NGC1893} 
\maketitle
\section{Introduction}
A key issue of the star formation process is its dependence on environmental conditions. Theoretically, temporal and spatial variations are expected since, for example, a fundamental parameter of star formation, the Jeans mass, depends on gas temperature, chemical abundances, and density \citep[cf.][]{Elmegreen_02, Elmegreen_04}. Moreover the variation in the initial mass function at low stellar masses across different environments suggests that cloud fragmentation based on the Jeans formulation, i.e., on gravity, is not the only mechanism at work, but that other processes (e.g. turbulence) are also important \citep{Padoan_99,Maclow_04}.

We observe stars forming under very different conditions and one of the key questions is whether star formation proceeds in the same way under the very different environmental conditions of the inner and outer region of the Galaxy. Molecular clouds and very young stellar associations are also present at large distances from the Galactic Center \citep{Snell_02}, implying that star formation still occurs in these regions. Conditions in the outer Galaxy should be much less conducive to star formation: the average surface and volume densities of atomic and molecular hydrogen are much smaller than in the solar neighborhood or in the inner Galaxy \citep{Wouterloot_90}, the interstellar radiation field is weaker \citep{Mathis_83}, prominent spiral arms are lacking and there are fewer supernovae to act as external triggers of star formation. Metal content is, on average, smaller \citep{Wilson_92}, decreasing radiative losses and therefore increasing cloud temperatures and consequently pressure support. Moreover, the pressure of the inter-cloud medium is smaller \citep{Elmegreen_89}. Given these relevant differences, quantitative observations of star forming regions in the outer Galaxy may help to establish the role of physical conditions in the star formation process.

In order to investigate the star formation processes in the outer Galaxy, studied the young cluster NGC1893, whose galactocentric distance is $\geq 11$ Kpc. Lying in the Aur OB2 association toward the Galactic anti-center, NGC1893 is associated with the HII region IC 410. It contains a group of early-type stars with some molecular clouds but only moderate extinction; \citet{Tapia_91} and \citet{Vallenari_99} estimated an age of 4 Myr from stars already on the Main Sequence, but the presence of two O5V stars indicates that the region should be younger than $\sim$3 Myr; \citet{Marco_02} found PMS B-type stars consistent with age $< 1$ Myr. These results suggest that star formation is still ongoing and hence there may be a range of ages for stars in the cluster.
 
Several authors performed UBV photometry of the cluster \citep{Cuffey_73, Moffat_74, Massey_95b}. \citet{Vallenari_99} obtained NIR photometry of the cluster, concluding that NGC1893 may have a rich pre-main sequence population. \citet{Tapia_91} determined an $A_V=1.68$ and a dereddened distance modulus of $(V-M_V)_0=13.18 \pm0.11$, corresponding to a heliocentric distance of $4.3$ Kpc at $l=173.6 \degr$; other authors calculated the distance modulus corresponding to a distance in a range between $3.2$ and $6$ Kpc \citep{Massey_95b,Marco_01,Daflon_04,Sharma_07}. With the solar galactocentric distance taken to be $8.5$ Kpc, NGC1893 is therefore located at a distance of $11.7-14.5$ Kpc from the Galactic Center. NGC1893 has a slightly sub-solar metal content, as reported by \citet{Rolleston_00} from the analysis of spectra of a few B stars. \citet{Daflon_04} confirmed that the abundances of C, N, and O in NGC1893 were two to three times lower than in the Sun.

We present the first results of the joint \textit{Chandra-Spitzer} large project \textit{The Initial Mass Function in the Outer Galaxy: the star forming region NGC1893} aimed at studying star formation processes at high galactocentric distances using multi-wavelength data, including infrared, X-ray, and optical observations. By means of the infrared observations, we study the properties of members with disks and/or envelopes, while the analysis of X-ray emission is crucial to establishing the membership of stars without infrared excesses and is also the elective way to investigate the coronal properties. By means of the optical observations, we finally estimate the masses of the stars and determine the distance and the age of the cluster. The project, led by G. Micela, includes a $\sim 31.2\arcmin \times 26.0\arcmin$ IRAC/\textit{Spitzer} map of the cluster and 5 \textit{Chandra} ACIS observations with a total exposure time of $\sim 440$ Ks.\\
In this first work, we conduct a census of the pre-main sequence candidate members of NGC1893. The IRAC observations of NGC1893 allow us to determine the members of the cluster with infrared excesses: Class 0/I and Class II stars, while the 5 ACIS observations are used to select the Class III objects that have already lost their disks and have no prominent infrared excess.

In Sect. \ref{ir obs} we describe the IRAC observations and the data reduction. In Sect. \ref{chandra} we describe the X-ray observations, in Sect. \ref{membership} we identify the members of the cluster, while in Sect. \ref{results} we summarize our results. 

\section{Infrared Observations and Data Reduction}
\label{ir obs}
\subsection{IRAC Observations}
NGC1893 was observed with the Infrared Array Camera (\textit{IRAC}; \citeauthor{Fazio_04}, \citeyear{Fazio_04}) onboard the {\it Spitzer} Space Telescope on 2006 March 25. This instrument has four detector arrays, with each dedicated to a different wavelength band, centered at $3.6$, $4.5$, $5.8$ and $8.0$ $\mu$m, each of which has a field of view of $\sim5.2\arcmin \times 5.2\arcmin$. All four bands were observed simultaneously, using the High Dynamic Range (HDR) mode with a long integration frame of $26.8$ s and a short one of $1.0$ s. We covered the region of the cluster by a $5$ rows  $ \times \  5$ columns map with five dithers at each map position. The total observed area was $\sim 31.2\arcmin \times 26.0\arcmin$.

The data reduction started from the \textit{IRAC Basic Calibrated Data} (BCD), provided by the {\it Spitzer} Science Center (SSC) pipeline reduction (version S13.2.0). The SSC process includes flat field corrections, dark subtraction, linearity, and flux calibration. The pipeline also masks instrumental signatures, bad pixels, and saturated pixels due to bright sources.

We further processed the BCDs by means of the \textit{IRAC artifact correction} code, in order to mask bad pixels not identified by the SSC pipeline reduction, and correct muxbleed and column pull-down defects occurring close to moderately bright sources \footnote{All the documentation for this code is available at \textit{ http:// spider.ipac.caltech.edu/staff/carey/irac{\_}artifacts}}.

Using the \textit{SSC MOPEX} code \citep{Marleau_05}, we combined the BCDs into a single mosaic image for each channel and exposure time, by averaging together the overlapping frames; in this way, we obtained four images for the long integration frame time and four for the short one, with a S/N factor higher than for the individual BCDs.

\begin{figure*}[!th]
 \subfigure{\includegraphics[width=8cm]{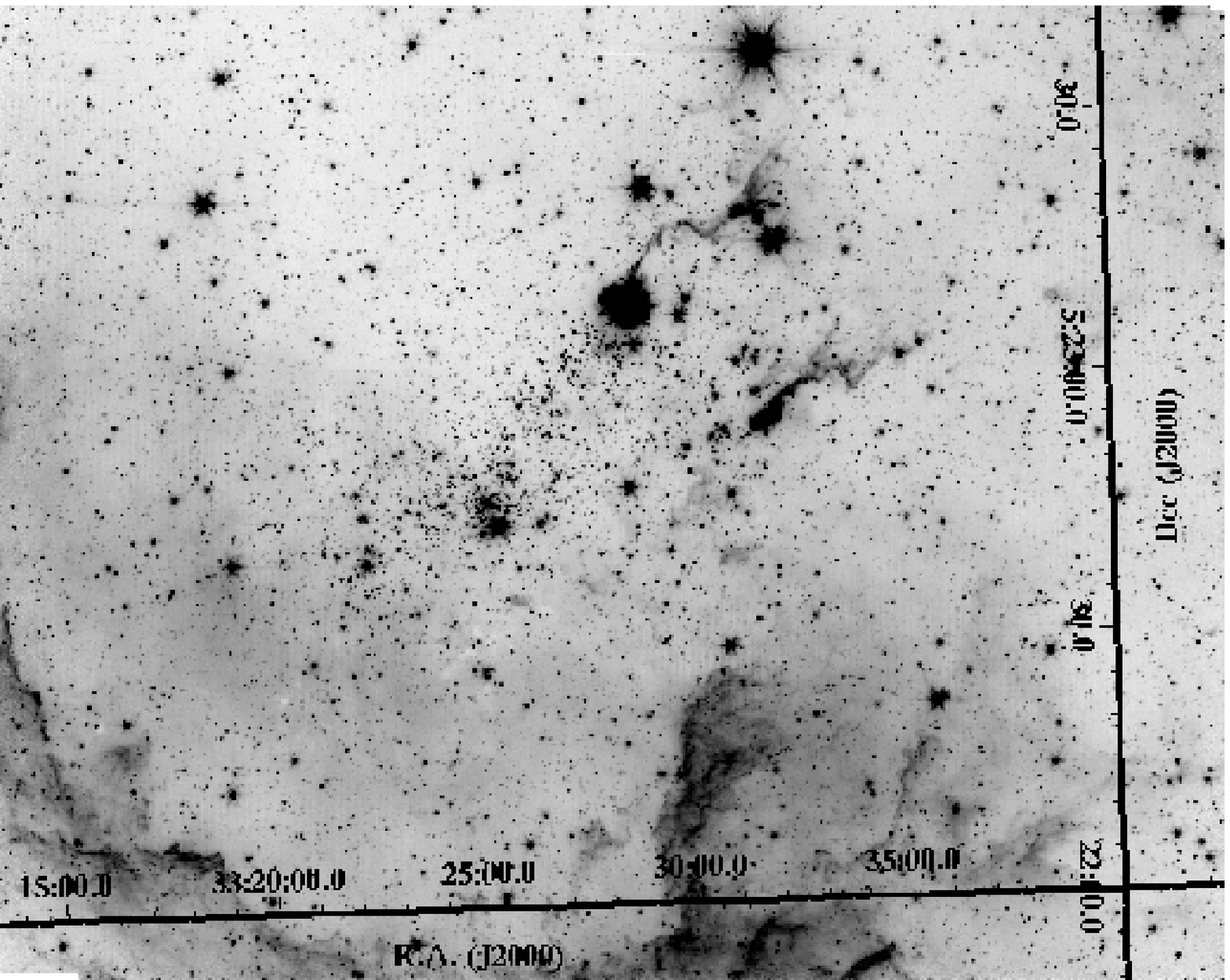}}
\hspace{5mm}
\subfigure{\includegraphics[width=8cm]{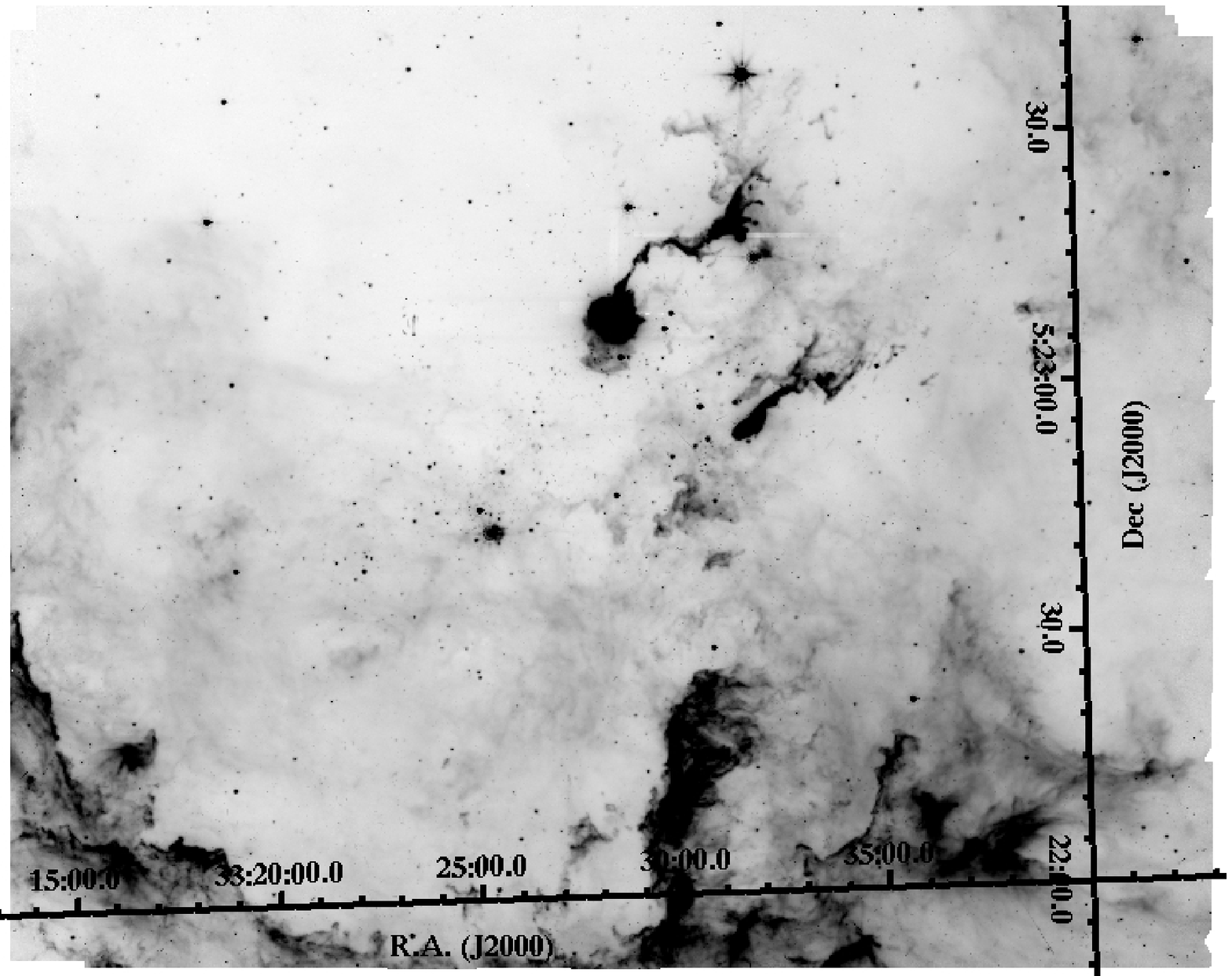}}\\
\caption{ Images of the $\sim 31.2\arcmin \times 26.0\arcmin$ mosaic for \textit{IRAC} $4.5$ (left) and $8.0$ $\mu$m (right) channels.}
\label{mosaic}
\end{figure*}

Figure \ref{mosaic} shows the mosaics obtained for the $4.5$ and $8.0$ $\mu$m channels. The figure illustrates that the two short-wavelength channels are far more sensitive for detecting stars than the two longer wavelength channels both because the detectors for the 3.6 and 4.5 $\mu$m channels are far more sensitive and because normal stellar photospheres are brighter at 4.5 than 8.0 microns.  The nebula associated with NGC1893, however, is far more evident at 8.0 $\mu$m because there are prominent PAH emission features in that band.

\subsection{Infrared Photometry and Catalogue}
\label{photometry}
We analyzed each mosaic image with the \textit{DAOPHOT II} procedures \citep{Stetson_87}, to obtain a list of point sources, their positions, and their magnitudes. Since \textit{DAOPHOT} expects images calibrated in electrons (or electrons/second), we converted the surface flux images back into counts, using the flux conversion values found in the \textit{IRAC Data Handbook}\footnote{\textit{ http:// ssc.spitzer.caltech.edu/irac/dh/iracdatahandbook3.0.pdf}}. We found star-like objects using a detection threshold of $5 \sigma$ for $3.6$ and $4.5$ $\mu$m channels and $6 \sigma$ for $5.8$ and $8.0$ $\mu$m channels, where $\sigma$ is the standard deviation of sky counts \footnote{The detection threshold is higher in 5.8 and 8.0 $\mu$m channels to avoid spurious sources corresponding to the nebula peaks}.

We derived aperture photometry with an aperture radius of $3$ pixels ($3.6\arcsec$), considering  an annulus with inner radius $12\arcsec$ and outer radius $24\arcsec$ for the background extraction, and we applied the relevant aperture corrections. For the zero point fluxes and the aperture corrections, we used values found in the \textit{IRAC Data Handbook} \citep[see also][]{Reach_05}. Using the \textit{DAOPHOT II/ALLSTAR} \citep{Stetson_87} and \textit{ALLFRAME} \citep{Stetson_94} procedures, we also computed the PSF photometry, calibrating the PSF magnitudes by means of a comparison with the aperture photometry.
Since, for the bright stars, the uncertainties related to the PSF magnitudes are larger than the aperture ones, and, on the other hand, the PSF errors are small for faint stars, we decided to use, for each star the magnitude computed with the procedure providing the smallest uncertainty.

In order to eliminate false sources, we merged the four IRAC source lists and the 2MASS K band catalogue, by cross-identifying sources, using a radius of $0.6\arcsec$. We accepted a star in the final catalogue if we were able to identify it in at least two different wavelengths. Moreover we considered only the magnitudes that have an uncertainty of less than $0.1$. In Table \ref{table_c}, we provide our catalogue, which includes the sequential numbers, the celestial coordinates, the magnitudes in K, and in  $3.6$, $4.5$, $5.8$, and  $8.0$ $\mu$m and the DAOPHOT {\tt sharp} parameter, that measures how well the brightness distribution could be interpreted as a point-like function (we use this parameter in the following to identify extended sources).

In Table \ref{num}, we report the total number of detected objects in the catalogue, specifying the number of sources in each \textit{IRAC} channel and in the 2MASS catalogue and the detection limit for each band. At $5.8\ \mu$m and  $8.0$ $\mu$m, where the instrument sensitivity is lower than in the other two channels and the background is higher, we find fewer sources. Although the number of sources common to the four channels is strongly limited by the sensitivity of these two channels, we found 1028 objects, 25\% of which were not previously detected with 2MASS.  
\begin{sidewaystable*}
\vspace{2.5truecm}
\tabcolsep 0.15truecm
\caption{Data Catalogue. $ID$ is the
source identification number, $RA(J2000)$ and $DEC(J2000)$ are the celestial coordinates, $K$ is the 2MASS K magnitude and $\sigma_K$
is its photometric uncertainty,
 $[3.6]$, $[4.5]$, $[5.8]$, $[8.0]$ are the magnitudes in the four
 IRAC channels and $\sigma_{[3.6]}$
, $\sigma_{[4.5]}$, $\sigma_{[5.8]}$, $\sigma_{[8.0]}$ are their photometric uncertainties. The complete table is available in electronic format at the CDS.} 
\centering
\begin{tabular}{cccccccccccccc}

\hline
\hline
\\
$ID$ & $RA(J2000)$ & $DEC(J2000)$ & $K$ &$\sigma_K$& $[3.6]$& $\sigma_{[3.6]}$&$[4.5]$& $\sigma_{[4.5]}$ &$[5.8]$&
$\sigma_{[5.8]}$ &$[8.0]$& $\sigma_{[8.0]}$ & {\tt sharp}\\
\hline
$26$ &$5:21:51.59$ &$33:30:43.03$ &$10.62$ &$0.02$ &$-$ &$-$ &$10.492$ &$0.003$ &$11.006$ &$0.095$ &$10.603$ &$0.071$&$-0.486$\\
$91$ &$5:21:52.94$ &$33:21:49.24$ &$10.60$ &$0.02$ &$-$ &$-$ &$10.506$ &$0.002$ &$10.556$ &$0.048$ &$10.389$ &$0.043$&$0.065$\\
$121$ &$5:21:53.43$ &$33:27: 9.55$ &$12.31$ &$0.02$ &$11.367$ &$0.003$ &$10.865$ &$0.004$ &$10.230$ &$0.004$ &$9.700$ &$0.008$&$-0.001$\\
$142$ &$5:21:53.74$ &$33:23:26.93$ &$13.01$ &$0.03$ &$12.777$ &$0.013$ &$12.748$ &$0.011$ &$12.041$ &$0.081$ &$-$ &$-$&$0.234$\\
$151$ &$5:21:53.87$ &$33:26: 8.63$ &$9.64$ &$0.02$ &$9.441$ &$0.046$ &$9.397$ &$0.081$ &$9.413$ &$0.004$ &$9.378$ &$0.016$&$-0.027$\\
$159$ &$5:21:53.96$ &$33:34:55.75$ &$12.60$ &$0.02$ &$12.391$ &$0.006$ &$12.424$ &$0.008$ &$12.421$ &$0.039$ &$12.592$ &$0.098$&$-0.082$\\
$166$ &$5:21:54.02$ &$33:31:42.45$ &$13.58$ &$0.03$ &$13.685$ &$0.031$ &$13.629$ &$0.031$ &$13.609$ &$0.041$ &$-$ &$-$&$0.015$\\
$176$ &$5:21:54.15$ &$33:34:24.26$ &$-$ &$-$ &$13.811$ &$0.069$ &$-$ &$-$ &$11.413$ &$0.077$ &$9.602$ &$0.090$&$0.664$\\
$183$ &$5:21:54.26$ &$33:35: 6.90$ &$14.40$ &$0.06$ &$14.691$ &$0.042$ &$14.298$ &$0.034$ &$14.069$ &$0.091$ &$-$ &$-$&$0.185$\\
$189$ &$5:21:54.30$ &$33:27:42.04$ &$13.97$ &$0.05$ &$13.699$ &$0.028$ &$13.742$ &$0.036$ &$13.513$ &$0.058$ &$-$ &$-$&$0.160$\\
$196$ &$5:21:54.38$ &$33:27:34.23$ &$-$ &$-$ &$15.924$ &$0.044$ &$15.082$ &$0.057$ &$14.164$ &$0.082$ &$-$ &$-$&$0.067$\\
$197$ &$5:21:54.43$ &$33:32: 0.52$ &$12.36$ &$0.02$ &$12.326$ &$0.007$ &$12.290$ &$0.008$ &$12.046$ &$0.041$ &$11.280$ &$0.079$&$0.156$\\
$236$ &$5:21:54.85$ &$33:27: 5.93$ &$14.71$ &$0.08$ &$14.499$ &$0.029$ &$14.523$ &$0.033$ &$14.069$ &$0.085$ &$-$ &$-$&$0.138$\\
$238$ &$5:21:54.89$ &$33:33: 9.28$ &$12.60$ &$0.02$ &$12.413$ &$0.005$ &$12.530$ &$0.006$ &$12.452$ &$0.028$ &$12.362$ &$0.094$&$0.022$\\
$246$ &$5:21:55.00$ &$33:30:50.75$ &$11.87$ &$0.02$ &$11.682$ &$0.011$ &$11.827$ &$0.013$ &$11.860$ &$0.063$ &$-$ &$-$&$-0.109$\\
$297$ &$5:21:55.66$ &$33:29:23.51$ &$-$ &$-$ &$13.684$ &$0.083$ &$-$ &$-$ &$11.179$ &$0.093$ &$9.277$ &$0.092$&$0.576$\\
$301$ &$5:21:55.75$ &$33:22:42.11$ &$12.24$ &$0.02$ &$12.206$ &$0.009$ &$12.068$ &$0.008$ &$12.189$ &$0.025$ &$12.106$ &$0.043$&$0.115$\\
$302$ &$5:21:55.76$ &$33:33:52.78$ &$12.22$ &$0.02$ &$11.903$ &$0.005$ &$11.985$ &$0.007$ &$11.950$ &$0.026$ &$11.817$ &$0.097$&$0.060$\\
$303$ &$5:21:55.79$ &$33:26:34.29$ &$-$ &$-$ &$14.866$ &$0.031$ &$15.203$ &$0.081$ &$12.423$ &$0.044$ &$10.591$ &$0.035$&$0.776$\\
$305$ &$5:21:55.82$ &$33:35: 8.31$ &$13.71$ &$0.04$ &$13.699$ &$0.011$ &$13.773$ &$0.012$ &$13.635$ &$0.098$ &$13.159$ &$0.086$&$0.141$\\
$325$ &$5:21:56.05$ &$33:29:27.15$ &$-$ &$-$ &$13.597$ &$0.081$ &$-$ &$-$ &$11.138$ &$0.097$ &$9.245$ &$0.094$&$0.592$\\
...
\\
\hline
\hline
\label{table_c}
\end{tabular}
\end{sidewaystable*}
 
\begin{table*} 
\vspace{2.5truecm}
\tabcolsep 0.15truecm
\caption{ Number of infrared detected sources in the NGC1893 region and detection limits}
\centering
\begin{tabular}{ccc}

\hline
\hline
\\
list & Number of sources &Detection limit (mag)\\
\\
\hline
All catalog * & $15277$ & $-$\\
$ 3.6$ $\mu$m & $12176$ & $19.03$\\
$ 4.5$ $\mu$m & $10316$ & $18.34$\\
$ 5.8$ $\mu$m & $2932$ & $15.88$\\
$ 8.0$ $\mu$m & $1705$ & $14.92$\\
 K 2MASS  & $2203$ & $15.01$\\
 $4$ IRAC bands & $1028$ & $-$\\
$ 4$ IRAC bands and K 2MASS & $771$ & $-$\\
\hline
\hline
\multicolumn{3}{l} {*  Sources with at least two magnitudes and for which at least one magnitude has photometric error $<0.1$. }
\end{tabular}
\label{num}
\end{table*}

\section{X-ray Observations and Data Reduction}
\label{chandra}
\begin{table*}
\vspace{2.5truecm}
\tabcolsep 0.15truecm
\caption{X-ray \textit{Chandra} ACIS observations }
\centering
\begin{tabular}{cccccc}

\hline
\hline
\\
 ObsID & Start Time & Exposure & R.A. (J2000.0) & Decl. (J2000.0) & Roll Angle\\
      &          & (ks) & (deg) & (deg) & (deg)\\
\hline
8462 & 2006 Nov 07 13:33:15 & $42.61$ & $80.7088633$ & $33.4726607$ & $107.50840$\\
6406 & 2006 Nov 09 12:51:18 & $115.7$ & $80.7088706$ & $33.4726628$ & $107.50839$\\
6407 & 2006 Nov 15 05:31:19 & $126.2$ & $80.7088692$ & $33.4726604$ & $107.49471$\\
8476 & 2006 Nov 17 10:55:16 & $53.26$ & $80.7088737$ & $33.4726629$ & $107.49472$\\
6408 & 2007 Jan 23 00:12:30 & $102.8$ & $80.7054833$ & $33.4640840$ & $262.11412$\\
\hline
\hline
\end{tabular}
\label{chandra_obs}
\end{table*}
The X-ray observations of NGC1893 combine four nearly consecutive exposures of the cluster taken in 2006 November and a fifth exposure taken in 2007 January (see Table \ref{chandra_obs} of details), with a total exposure time of $\sim 440$ Ks. The detailed analysis of these X-ray data will be the subject of a future paper. Here we use the X-ray observations to identify probable members among the IRAC sources without any infrared excess in order to estimate the number of Weak-lined T Tauri stars. Therefore, we summarize only the main steps of the analysis.

The observations were obtained with the ACIS camera onboard \textit{Chandra} \citep{Weisskopf_02,Garmire_03} in the FAINT mode; we considered only results from ACIS-I, the four front-side illuminated CCDs covering about $17\arcmin \times 17\arcmin$ on the sky. ACIS-I is characterized by high angular resolution, with an on-axis PSF FWHM of $\sim0.5 \arcsec$, which makes it possible to resolve the central region of the cluster, despite the large distance of NGC1893.

We detected sources applying the PWDetect code \citep{Damiani_97} to the merged event file of the five observations, filtering photons with energy between $0.5$ and $8.0$ KeV. The significance threshold was set to $4.6$ $\sigma$, corresponding to 10 spurious sources detected in source-free fields with the background level of our observations. PWDetect reported 1025 sources. Upon careful inspection, we removed four entries corresponding to sources that were detected twice, leaving a total of 1021 distinct sources. 

\begin{figure*}[tb]
\centerline{\psfig{figure=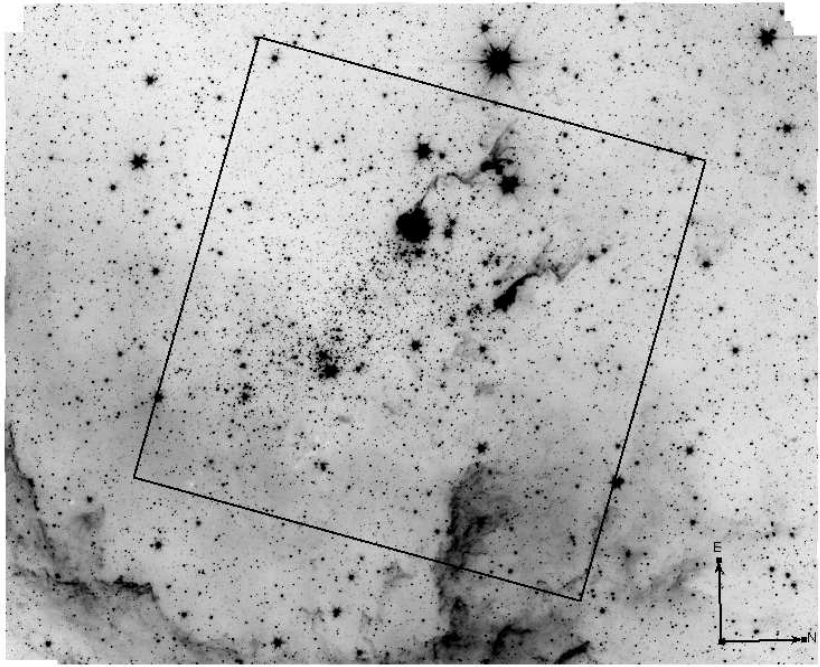,width=11 cm,angle=0}}
\caption{\textit{ACIS} field of view ($17\arcmin \times 17\arcmin$) on the \textit{IRAC} image at $4.5$ $\mu$m.}
\label{chandra_field}
\end{figure*}

In order to identify young stars in which the primordial disk had already dissipated, we combined the \textit{Spitzer} and \textit{Chandra} data. Since the field of view of \textit{Chandra} ACIS-I is limited to $17\arcmin \times 17\arcmin$, we were unable to find X-ray counterparts for all objects in the \textit{Spitzer} field of view. For comparison, we overplot the ACIS field of view on the IRAC image at $4.5 \mu$m (Fig. \ref{chandra_field}).

Due to the degrading of the ACIS PSF at high off axis, we used three different matching radii ($0.7\arcsec,1.5\arcsec,2.5\arcsec$), depending on the source off axis angle. The results of the match between the IR and the X-ray catalogues are given in Sect. \ref{weak_lined}.
\section{Young Stellar Objects in NGC1893}
\label{membership}
\subsection{IRAC four band source characterization}
In this section we analyze the subsample of sources detected in all four IRAC channels to find the likely members with excesses. We first reject the contaminants (extragalactic sources or nebular  emission) and then classify the young stars with excesses. 
\subsubsection{Estimate of Contaminants}
\label{extra}
\begin{figure*}[tb]
\centerline{\psfig{figure=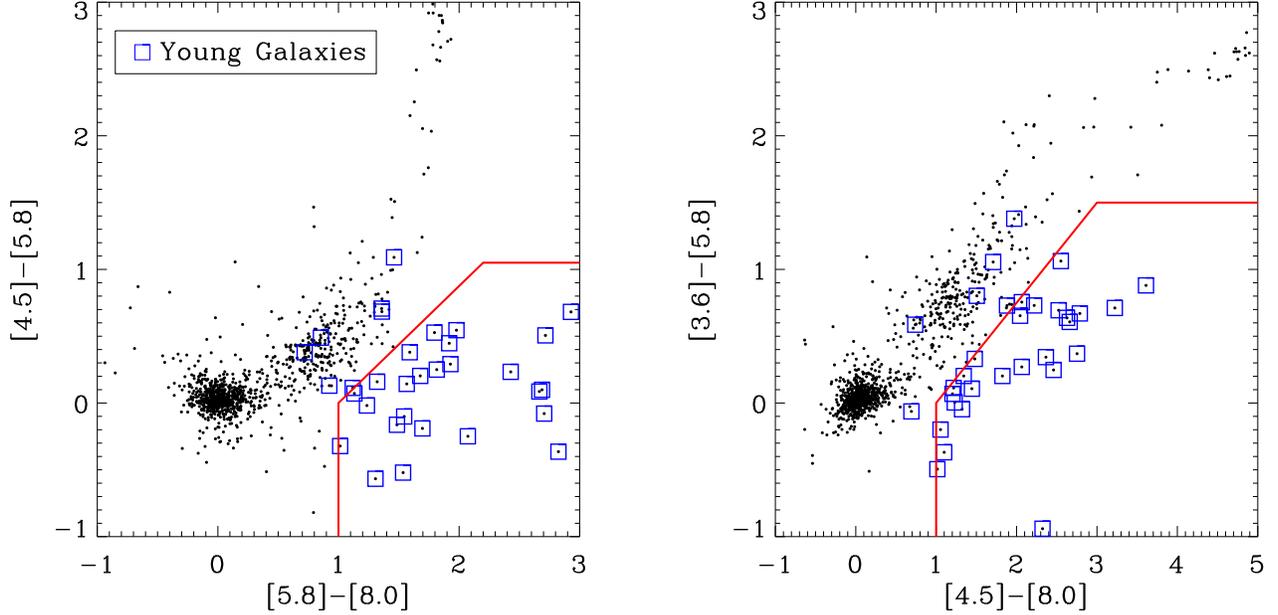,width=18 cm,angle=0}}
\caption{IRAC color-color diagrams illustrating the selection of star-forming galaxies, characterized by a strong PAH emission. Using the method described in \citet{Gutermuth_07}, we rejected the sources lying in the region delimited by the solid line in at least one of the two color-color diagrams. As explained in the legend, we designate the PAH-rich galaxies with little squares.}
\label{galaxy}
\end{figure*} 
\begin{figure}[tb]
\centerline{\psfig{figure=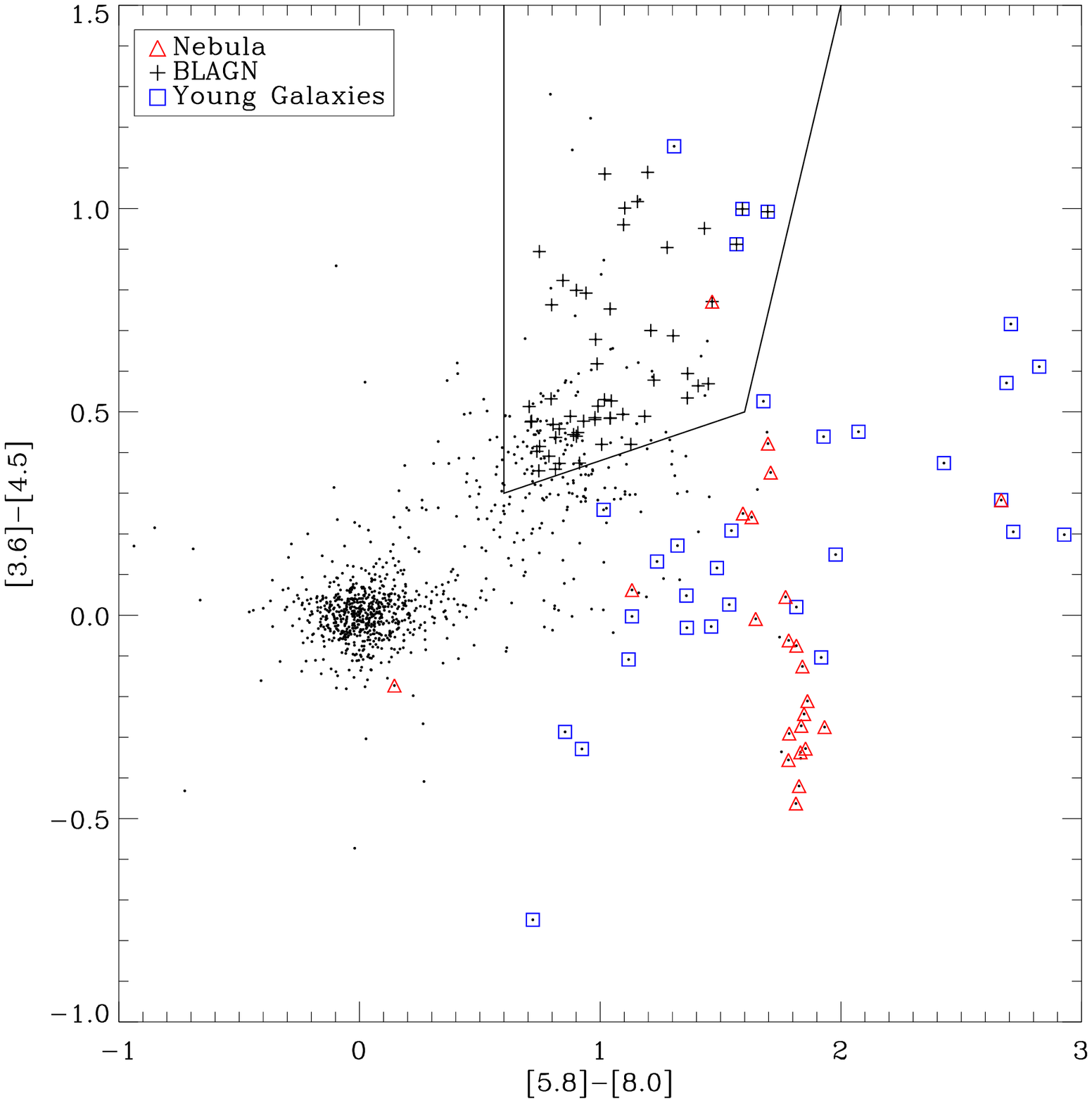,width=10 cm,angle=0}}
\caption{IRAC color-color diagram highlighting the contaminants to the young star sample. As explained in the legend, squares indicate the star-forming galaxies selected by means of the diagrams in Fig. \ref{galaxy}, crosses indicate the BLAGN, selected as the weak sources above the solid line, while triangles indicate the sources contaminated by the emission of the nebula: most of them are characterized by a strong red color in [5.8]-[8.0] and a blue color in [3.6]-[4.5].}
\label{agn}
\end{figure} 
Young stars with disks are characterized by infrared excesses, therefore the study of their Spectral Emission Distribution (SED) (either via their SED slope or their position in color-color diagrams) is the most powerful method for distinguishing the cluster members with excesses from the old stars in the same field of view. Nevertheless, there are at least two classes of contaminants: the extragalactic sources, mainly broad-line AGN (BLAGN) and star forming galaxies, and emission from the nebula that, for very embedded stars could contaminate the photometry.

To avoid contamination from extragalactic sources, several authors  \citep{Harvey_06,Jorgensen_06, Hernandez_07} in their study of nearby star forming regions, suggested a cut in magnitude at $3.6$ $\mu$m or at  $4.5$ $\mu$m. However, this method alone is less effective in the case of NGC1893  because it is much further from the Sun than the star forming regions considered in those studies, and its PMS stars are therefore faint.

Following \citet{Stern_05} and \citet{Gutermuth_07}, we decided to identify star-forming galaxies by means of their typical Polycyclic Aromatic Hydrocarbons (PAH) emission. Using the [4.5]-[5.8] vs. [5.8]-[8.0] and the [3.6]-[5.8] vs. [4.5]-[8.0] color-color diagrams in Fig. \ref{galaxy} and the equations given in \citet{Gutermuth_07}, it is possible to distinguish the regions delimited by the solid lines in Fig. \ref{galaxy}, where objects whose SED are dominated by PAH emission (indicated by empty squares) are typically found.

Since only a small number of young stars were shown to have significant PAH emission in other star forming regions \citep{Hartmann_05, Hernandez_07}, we rejected the 32 sources that satisfied at least one of the conditions described in \citet{Gutermuth_07}.

\citet{Stern_05} found broad-line AGNs in the following region of the [3.6]-[4.5] vs. [5.8]-[8.0] diagram:
\begin{center}
  \begin{equation} \label{eq: agn}
\begin{array}{l} 
\left[5.8\right]-\left[8.0\right]>0.6\\
\left[3.6\right]-\left[4.5\right]>0.2 \cdot \left( \left[5.8\right]-\left[8.0\right]\right) +0.18 \\
\left[3.6\right]-\left[4.5\right]>2.5 \cdot \left( \left[5.8\right]-\left[8.0\right]\right) -3.5 \\
\end{array}	 
\end{equation}
\end{center}
Due to the distance of NGC1893, a simple cut in magnitude would be ineffective to filter AGNs alone, and, on the other hand, the zone found by \citet{Stern_05} is similar to that occupied by protostars; we therefore decided to combine the two conditions, selecting AGNs as the sources that follow the conditions (\ref{eq: agn}) and whose $3.6$ $\mu$m magnitude is greater than 14.5 \citep{Hernandez_07}. Using this criterion, we identified 56 candidate AGN.

As sources contaminated by the nebular emission, we selected 23 objects with {\tt sharp} \citep{Stetson_87}  greater than $0.37$, corresponding to $3\sigma$ of the {\tt sharp} distribution , i.e. objects with a brightness distribution inconsistent with a point-like source. These sources have very red [5.8]-[8.0] colors, and most have a blue [3.6]-[4.5] color, due to the fact that the nebula is brighter at $4.5$ $\mu$m. It is possible that some of these objects are embedded members of the cluster and we do not classify them because their photometry could be contaminated by the nebular emission.

We illustrate the contaminants we identified by means of these criteria in Fig. \ref{agn}. Squares represent the candidate star-forming galaxies, found by means of diagrams in Fig. \ref{galaxy}, crosses (above the solid line) indicate candidate AGNs, and triangles represent sources contaminated by the emission of the nebula.
\subsubsection{Members with infrared excesses}
\label{excesses_members}
There are two standard ways to classify young stellar objects, both based on the analysis of their SED. The first method \citep{Lada_87} uses the shape of the SED, which in \textit{IRAC} bands can be approximated to a power law. From a comparison between models and observed SED, \citet{Lada_06} found that, in the \textit{IRAC} bands, stars with primordial disks have a slope $\alpha>-1.8$; stars with purely photospheric emission should have
 $\alpha<-2.56$; stars with slopes intermediate between these two values
are plausibly ``transitional objects'' that still have disks but possibly
with inner holes or with optically thin emission regions. Another standard method in the literature \citep{Allen_04, Megeath_04} consists of the conversion of SED models into position in the [3.6]-[4.5] versus [5.8]-[8.0] diagram.

Here we used a method similar to that of \citet{Allen_04} and \citet{Megeath_04}, with an extensive grid of models of young stars \citep{Robitaille_06}, to define regions in color-color diagrams relative to different PMS classes.

Following the \citet{Robitaille_06} notation for which $\dot{M}_{env}$ is the accretion rate from the envelope, ${M}_{\star}$ is the mass of the central star, and ${M}_{disk}$ is the mass of the disk, it is possible to define class 0/I stars as those with:
\begin{equation} \label{eq: proto}
\dot{M}_{env}/{M}_{\star} > 10^{-6}$ yr$^\mathrm{-1} \\
\end{equation} 
and class II stars as:
\begin{equation} \label{eq: disk}
\begin{array}{l}
\dot{M}_{env}/{M}_{\star} < 10^{-6}$ yr$^\mathrm{-1} \\
{M}_{disk}/{M}_{\star} >10^{-6} 
\end{array}
\end{equation} 
Converting the SEDs of these groups of models into IRAC colors  \citep[see figures 17 and 18 in][]{Robitaille_06}, we note that the class II objects occupy the region of the color-color diagram defined by $[3.6]-[4.5]<0.8$ and $[5.8]-[8.0]>0.4$ \footnote{We analyzed only models with inner disk radius less than 100 times the dust sublimation radius. Stars with internal disk radius greater than this limits have [3.6]-[4.5] and [5.8]-[8.0] colors close to zero, and the first color different from zero is [8.0]-[24.0], where we do not have observations.}.

\begin{figure}[tb]
\centerline{\psfig{figure=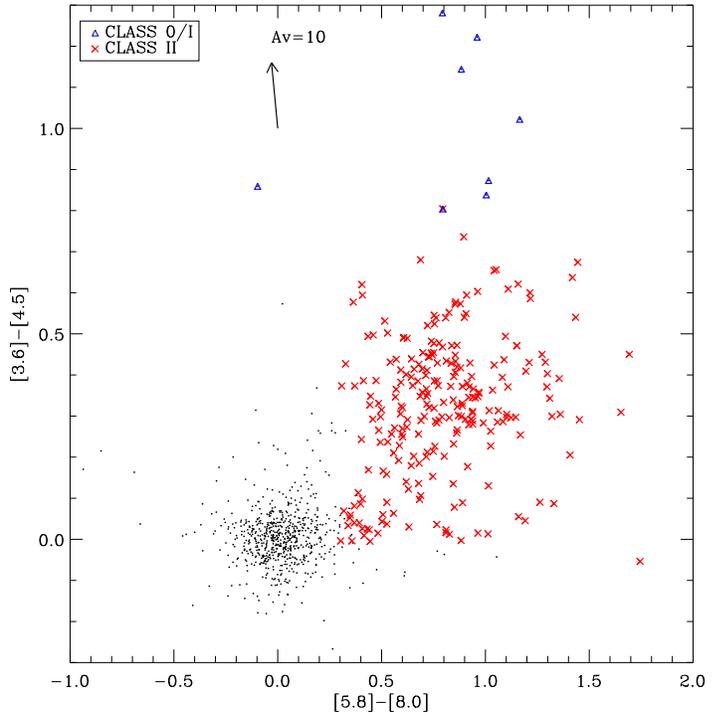,width=10 cm,angle=0}}
\caption{[5.8]-[8.0]vs.[3.6]-[4.5] IRAC color-color diagram. Triangles represent the objects whose SED is dominated by the envelope emission, while crosses refer to objects with SED dominated by the disk emission. The arrow refers to the reddening vector for $A_V=10$. To better classify the stars at the border between these two classes, we took into account the error associated with the colors.}
\label{classif}
\end{figure}

Analyzing the colors relative to class 0/I protostars, the models occupy a large region of the color-color diagram because of temperature and inclination effects. Although the locus $[3.6]-[4.5]>0.8$ is typical of class 0/I objects, the region occupied by class II stars can also be occupied by protostar models. Therefore, although many class 0/I stars can be uniquely identified, the evolutionary classifications of the remaining class 0/I as well as most class II objects cannot be derived from the IRAC colors alone. However, considering the different duration of the two stages,  in a cluster of ~3 Myrs it is ~10-100 times less probable to find a class 0/I than a class II star. Therefore, we designate as class 0/I only the 7 stars whose colors place them in the region uniquely corresponding to protostars, while we designate the 242 stars lying in the color-color diagram region common to the two evolutionary stages as class II objects. Using this method, we are able to identify 249 members of the cluster. In Fig. \ref{classif}, we plot the IRAC color-color diagram for our stars, using the previous classification. As explained in the legend, triangles represent the class 0/I objects, while the crosses indicate the class II stars.\\

\subsubsection{Weak lined T Tauri}
\label{weak_lined}
Although the star formation process is often considered a low energy process, high energy processes occur as early as the protostellar phase and pre-main sequence (PMS) stars are strong X-ray sources with luminosity up to $10^4$ times the X-ray luminosity of the present-day Sun. Among the sources detected by IRAC, we utilize this property to identify young stars that have already dissipated their disks  and would otherwise be indistinguishable from field stars (Class III or Weak lined T Tauri).

Since we are able to compare only the objects in the same field of view, we consider for this analysis the 12513 infrared sources in the $17\arcmin \times 17\arcmin$ ACIS-I/\textit{Chandra} FOV. Applying the analysis described in Sect. \ref{extra} to the 692 sources with emission in all four IRAC channels, we labeled 77 sources as possible contaminants. As discussed in Sect. \ref{excesses_members}, by means of the IRAC color-color diagram we identified 219 stars with disks and 6 objects with envelope emission.\\
In our \textit{ACIS} observations, we identify 1021 X-ray sources. We expect that this sample is strongly contaminated by extragalactic sources because NGC1893 is located towards the anticenter direction. Using the galactic H column density $N_H \sim 6 \times 10^{21}$ cm$^\mathrm{-2}$ \footnote{calculated by means of the $N_H$ tools of \textit{HEASOFT} (see http://heasarc.gsfc.nasa.gov/cgi-bin/Tools/w3nh/w3nh.pl)}, the flux limit of our X-ray observations and the results from the \textit{Chandra North Deep Field} \citep {Brandt_01}, we predict 300-550 extragalactic contaminants in our X-ray catalogue.

By  matching the X-ray catalogue with all the 12513 infrared sources in our catalogue of sources in the ACIS field of view, we find 720 matches ($\sim70 \%$ of the X-ray sources). Considering the 692 infrared sources detected in all the four IRAC energy bands, we find 248 matches, 20 of which correspond to extragalactic contaminants and one to a source contaminated by the emission of the nebula, according to the criteria discussed in Sect. \ref{extra}; we therefore consider the remaining 227 sources as candidate members of the cluster.

\begin{figure}[tb]
\centering
\includegraphics[width=9cm]{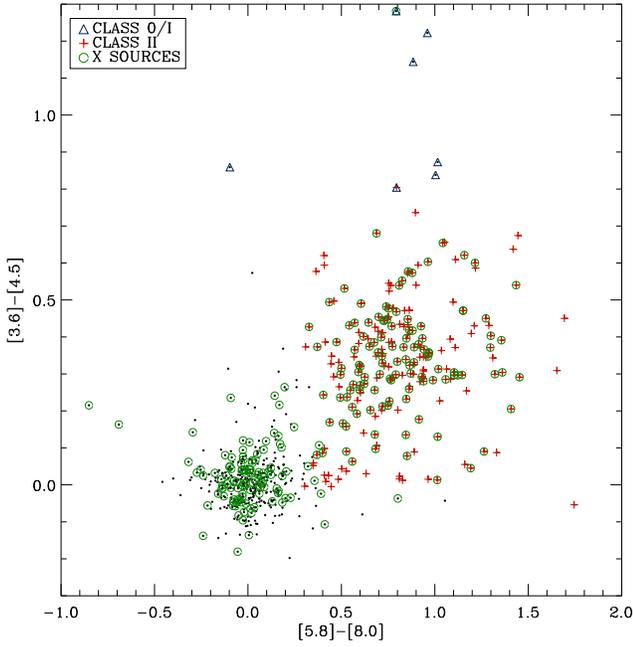}
\caption{IRAC color-color diagram for stars in the \textit{Chandra} ACIS-I field of view. Dots represent the infrared sources and circles mark IRAC sources with X-ray counterparts. We overplot the classification of objects dominated by envelope emission (triangles) and stars dominated by disk emission (crosses).}
\label{match_x_classif}
\end{figure}

In Fig.\ref{match_x_classif}, we report the color-color diagram of objects in the \textit{ACIS} field of view (without infrared candidate extragalactic sources): among the 227 candidate members with X-ray emission, we distinguish 116 stars, whose infrared emission is dominated by disk, 1 object with a strong envelope emission and 110 sources in the color-color region corresponding to photospheres of normal stars. We consider these 110 sources to be Class III members of the cluster.

\subsection{Additional YSOs identified via other diagrams}
\label{other_members}
In the previous sections, we limited our analysis to the stars detected in all four IRAC bands, to find the contaminants and classify the pre-main sequence stars in our field of view. Nevertheless, the above selection criterion strongly limits the number of objects, mainly because of the lower sensitivity of the $5.8$ and $8.0$ $\mu$m channels: we found in these channels $\sim 10 \%$ of the detections found in the $3.6$ or $4.5$ $\mu$m channels (cf Table \ref{num}).

\begin{figure}[tb]
\centerline{\psfig{figure=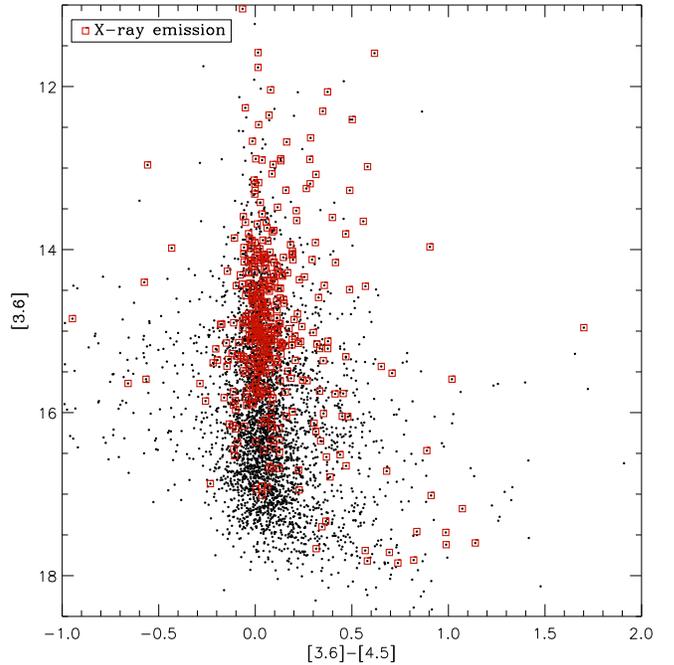,width=9.3 cm,angle=0}}
\caption{Color-magnitude diagram for stars detected in the $3.6$ and $4.5$ $\mu$m bands but not in all four IRAC bands. The squares represent objects with X-ray emission. The 459 stars with $[3.6]-[4.5] > 0.3$ and {\tt sharp} parameter compatible with a point-like object are candidate members of NGC1893 with infrared excesses.}

\label{candi}
\end{figure}

Considering the 6768 stars detected in the $3.6$ and $4.5$ $\mu$m channels that are not detected in all the four IRAC bands, we found 704 stars that have color $[3.6]-[4.5] > 0.3$ and {\tt sharp} value compatible with that of a point-like source. In the ACIS field of view, where there are 3890 infrared detections and 408 stars with X-ray emission, we found 459 stars with $[3.6]-[4.5]$ color excesses, 49 of which present X-ray emission. Because of their color typical of PMS stars with disks, these 459 stars are candidate members of the cluster, but we are unable to distinguish between protostars and Classical T-Tauri stars because of the lack of IRAC color-color information. In Fig. \ref{candi}, we plot $3.6$ $\mu$m magnitude versus $[3.6]-[4.5]$ color for stars detected in the ACIS field of view that have emission in the $3.6$ and $4.5$ $\mu$m channels but were not detected in all four IRAC channels, and overplot stars with X-ray emission. The main problem related to these objects is that we are unable to distinguish them from extragalactic contaminants. For this aim, optical spectroscopy could be useful to find young objects by means of measures of lithium abundance and $H_\alpha$ and Ca II lines (or, more basically, to distinguish between young stars at zero redshift, and AGN at redshifts up to a few tenths).

 Other authors \citep{Gutermuth_07,Lucas_07} use also the $[[K]-[3.6]]_0$ versus $[[3.6]-[4.5]]_0$  color-color diagram as a means of secondary identification of sources that lack $[5.8]$ or $[8.0]\ \mu$m photometric data; due to the sensitivity limit of 2MASS and distance of NGC1893, in our sample there are, however, only 10 stars of the previous 459 candidate members that show emission also in the K band, thus for our purposes these alternative methods are not useful for identifying new YSOs.

In Table \ref{member_cat}, we report all the candidate members of NGC1893; in the first five columns we report the ID and the IRAC magnitudes, while in the sixth and seventh columns we report two flags, the first indicating if an objects is a Class 0/I, a Class II or a Class III star, the second if a source is in the ACIS field and if it is an X-ray source.
\begin{table}
\vspace{2.5truecm}
\tabcolsep 0.15truecm
 \centering
\caption{Candidate Members of NGC1893. $ID$ is the
source identification number (as in Table \ref{table_c}), $[3.6]$, $[4.5]$, $[5.8]$, $[8.0]$ are the IRAC magnitudes, Class is a flag identifying the class of the source (0 for Class 0/I protostars, 2 for Classical T-Tauri stars and 3 for Weak-lined T-Tauri), where applicable, X-ray flag is -1 for stars out of the ACIS field of view, 1 for X-ray sources, 0 for stars without X-ray emission in the ACIS field of view}. The complete list of sources is available in electronic format.
\begin{tabular}{ccccccc}

\hline
\hline
\\

$ID$ & $[3.6]$& $[4.5]$& $[5.8]$&$[8.0]$& Class &  X-ray flag\\
\hline
$121$ &$11.37$ &$10.86$ &$10.23$ &$9.70$ &$2$ &$-1$\\
$197$ &$12.33$ &$12.29$ &$12.05$ &$11.28$ &$2$ &$-1$\\
$1013$ &$11.53$ &$11.41$ &$11.32$ &$10.69$ &$2$ &$-1$\\
$1264$ &$17.49$ &$17.19$ &$-$ &$-$ &$-$ &$0$\\
$1269$ &$17.58$ &$17.25$ &$-$ &$-$ &$-$ &$0$\\
$1305$ &$13.66$ &$13.62$ &$13.66$ &$13.32$ &$2$ &$-1$\\
$1439$ &$14.31$ &$13.82$ &$13.50$ &$12.88$ &$2$ &$-1$\\
$1592$ &$17.78$ &$16.78$ &$-$ &$-$ &$-$ &$0$\\
$1648$ &$15.67$ &$14.43$ &$-$ &$-$ &$-$ &$0$\\
$1716$ &$10.22$ &$10.23$ &$10.14$ &$9.78$ &$2$ &$-1$\\
$1718$ &$13.53$ &$13.47$ &$13.49$ &$12.98$ &$2$ &$-1$\\
$1770$ &$13.14$ &$12.62$ &$12.25$ &$11.53$ &$2$ &$-1$\\
$1888$ &$16.79$ &$16.49$ &$-$ &$-$ &$-$ &$0$\\
$1933$ &$17.63$ &$17.26$ &$-$ &$-$ &$-$ &$0$\\
$1939$ &$12.24$ &$12.20$ &$12.04$ &$11.41$ &$2$ &$0$\\
$1989$ &$17.72$ &$17.02$ &$-$ &$-$ &$-$ &$1$\\
$1994$ &$17.63$ &$17.19$ &$-$ &$-$ &$-$ &$0$\\
$2063$ &$13.48$ &$12.93$ &$12.40$ &$11.49$ &$2$ &$-1$\\
$2079$ &$10.51$ &$10.45$ &$10.50$ &$10.54$ &$3$ &$1$\\
$2105$ &$13.83$ &$13.79$ &$13.93$ &$13.56$ &$2$ &$-1$\\
\\
\hline
\hline

\end{tabular}
\label{member_cat}
\end{table}
 
\section{Summary and Conclusion}
\label{results}
We have analyzed a set of \textit{IRAC} observations of NGC1893, a young cluster, in the outer part of the Galaxy.
\begin{itemize} 
\item By analyzing the \textit{IRAC} images, we were able to identify in the field of view ($\sim 31.2\arcmin \times 26.0\arcmin$) 1028 sources emitting in all four \textit{IRAC} bands, 25\% of which have not been previously detected by 2MASS.
\item By analyzing the color-color diagram, we identified 242 stars with disks, and 7 Class 0/I protostars.  Moreover in the ACIS-I \textit{Chandra} field of view, we detected 110 Weak lined T Tauri. At this stage, the census of the cluster is incomplete: we have found 459 objects in the IRAC field that showed an infrared excess ($[3.6]-[4.5] > 0.3$) and were not detected at $5.8$ and $8.0$ $\mu$m. These sources are candidate members of the cluster, but we cannot distinguish them from contaminants. Moreover, we found a sample of 472 X-ray sources with no infrared counterpart, which is probably dominated by extragalactic sources.
\item Although the catalogue of members is incomplete, we have detected 359 members, indicating that NGC1893 is quite rich, with intense star forming activity despite the ``unfavorable'' environmental conditions in the outer Galaxy. 
\end{itemize} 
The near and mid-infrared data are unsuitable for estimating age and masses of the stars, because the presence of disks affects the magnitudes. To place our survey in context with other IRAC surveys of star forming regions, it is, however, useful to derive a rough estimate of the mass range for our census of NGC1893.  To achieve this, we assume that the J magnitude provided by 2MASS is entirely photospheric, for our Class III stars. Our brightest Class III star has $\rm{J} \sim 8.28$. In order to compare the J luminosity with models, we need to obtain the absolute magnitude value that depends on the supposed distance and absorption. Using distance and absorption measured from \citet{Tapia_91}, \citet{Marco_01}, and \citet{Sharma_07}, we converted the J magnitudes in absolute magnitudes. Comparing our calculated values with that obtained from models by \citet{Marigo_08}, we found that the highest mass is in the range $\sim 28-46 M_{\odot}$. To estimate the lower mass limit, we considered that our faintest Class III has $[3.6]=14.6$, and converted it to absolute magnitude in the same way as for the J magnitude. Comparing our value with the L magnitudes of PMS models by \citet{Siess_00} relative to the ages calculated by previous authors ($1-4$ Myr), we estimated that the lowest mass weak-lined T-Tauri star is in the range $\sim 1.4-3 M_{\odot}$.

 We found that the 3\% of our stars with IR excesses are Class 0/I protostars. This value is similar to the 2\% value found in IC348 \citep{Lada_06}, implying that our cluster does not show any peculiarity regarding the ongoing star formation, if compared with clusters of the same age in the inner Galaxy.

The fraction of stars with primordial disks that we measure in NGC1893 is about 67\%.  This value is an upper limit because our selection of diskless stars is based on the IR observations, therefore, we are able to detect stars with IR excesses to lower mass limits than for stars without disks. Considering that clusters of similar ages ($\sim 3-4$ Myr) may show fractions of disks in the range $\sim 50-61 \%$ with significant uncertainty \citep{Lada_06,Luhman_06}, our upper limit compares well with disk fractions measured in the nearby SFRs.
\begin{acknowledgements}
We acknowledge financial contribution from contract ASI-INAF I/023/05/0 and from European Commission (contract N. MRTN-CT-2006-035890).\\
This work is based on observations made with the Spitzer Space
Telescope, which is operated by the Jet Propulsion Laboratory, California
Institute of Technology under a contract with NASA. Support for this work
was provided by NASA through an award issued by JPL/Caltech.
\end{acknowledgements}

\bibliographystyle{aa}
\bibliography{0132}

\begin{thebibliography}{43}
\expandafter\ifx\csname natexlab\endcsname\relax\def\natexlab#1{#1}\fi

\bibitem[{{Allen} {et~al.}(2004){Allen}, {Calvet}, {D'Alessio}, {Merin},
  {Hartmann}, {Megeath}, {Gutermuth}, {Muzerolle}, {Pipher}, {Myers}, \&
  {Fazio}}]{Allen_04}
{Allen}, L.~E., {Calvet}, N., {D'Alessio}, P., {et~al.} 2004, \apjs, 154, 363

\bibitem[{{Brandt} {et~al.}(2001){Brandt}, {Alexander}, {Hornschemeier},
  {Garmire}, {Schneider}, {Barger}, {Bauer}, {Broos}, {Cowie}, {Townsley},
  {Burrows}, {Chartas}, {Feigelson}, {Griffiths}, {Nousek}, \&
  {Sargent}}]{Brandt_01}
{Brandt}, W.~N., {Alexander}, D.~M., {Hornschemeier}, A.~E., {et~al.} 2001,
  \aj, 122, 2810

\bibitem[{{Cuffey}(1973)}]{Cuffey_73}
{Cuffey}, J. 1973, \aj, 78, 408

\bibitem[{{Daflon} \& {Cunha}(2004)}]{Daflon_04}
{Daflon}, S. \& {Cunha}, K. 2004, \apj, 617, 1115

\bibitem[{{Damiani} {et~al.}(1997){Damiani}, {Maggio}, {Micela}, \&
  {Sciortino}}]{Damiani_97}
{Damiani}, F., {Maggio}, A., {Micela}, G., \& {Sciortino}, S. 1997, \apj, 483,
  370

\bibitem[{{Elmegreen}(1989)}]{Elmegreen_89}
{Elmegreen}, B.~G. 1989, \apj, 338, 178

\bibitem[{{Elmegreen}(2002)}]{Elmegreen_02}
{Elmegreen}, B.~G. 2002, \apj, 577, 206

\bibitem[{{Elmegreen}(2004)}]{Elmegreen_04}
{Elmegreen}, B.~G. 2004, Memorie della Societa Astronomica Italiana, 75, 362

\bibitem[{{Fazio} {et~al.}(2004){Fazio}, {Hora}, {Allen}, {et~al.}}]{Fazio_04}
{Fazio}, G.~G., {Hora}, J.~L., {Allen}, L.~E., {et~al.} 2004, \apjs, 154, 10

\bibitem[{{Garmire} {et~al.}(2003){Garmire}, {Bautz}, {Ford}, {Nousek}, \&
  {Ricker}}]{Garmire_03}
{Garmire}, G.~P., {Bautz}, M.~W., {Ford}, P.~G., {Nousek}, J.~A., \& {Ricker},
  Jr., G.~R. 2003, in Presented at the Society of Photo-Optical Instrumentation
  Engineers (SPIE) Conference, Vol. 4851, X-Ray and Gamma-Ray Telescopes and
  Instruments for Astronomy. Edited by Joachim E. Truemper, Harvey D.
  Tananbaum. Proceedings of the SPIE, Volume 4851, pp. 28-44 (2003)., ed. J.~E.
  {Truemper} \& H.~D. {Tananbaum}, 28--44

\bibitem[{{Gutermuth} {et~al.}(2007){Gutermuth}, {Myers}, {Megeath}, {Allen},
  {Pipher}, {Muzerolle}, {Porras}, {Winston}, \& {Fazio}}]{Gutermuth_07}
{Gutermuth}, R.~A., {Myers}, P.~C., {Megeath}, S.~T., {et~al.} 2007, ArXiv
  e-prints, 710

\bibitem[{{Hartmann} {et~al.}(2005){Hartmann}, {Megeath}, {Allen}, {Luhman},
  {Calvet}, {D'Alessio}, {Franco-Hernandez}, \& {Fazio}}]{Hartmann_05}
{Hartmann}, L., {Megeath}, S.~T., {Allen}, L., {et~al.} 2005, \apj, 629, 881

\bibitem[{{Harvey} {et~al.}(2006){Harvey}, {Chapman}, {Lai}, {Evans}, {Allen},
  {J{\o}rgensen}, {Mundy}, {Huard}, {Porras}, {Cieza}, {Myers}, {Mer{\'{\i}}n},
  {van Dishoeck}, {Young}, {Spiesman}, {Blake}, {Koerner}, {Padgett},
  {Sargent}, \& {Stapelfeldt}}]{Harvey_06}
{Harvey}, P.~M., {Chapman}, N., {Lai}, S.-P., {et~al.} 2006, \apj, 644, 307

\bibitem[{{Hern{\'a}ndez} {et~al.}(2007){Hern{\'a}ndez}, {Hartmann}, {Megeath},
  {Gutermuth}, {Muzerolle}, {Calvet}, {Vivas}, {Brice{\~n}o}, {Allen},
  {Stauffer}, {Young}, \& {Fazio}}]{Hernandez_07}
{Hern{\'a}ndez}, J., {Hartmann}, L., {Megeath}, T., {et~al.} 2007, \apj, 662,
  1067

\bibitem[{{J{\o}rgensen} {et~al.}(2006){J{\o}rgensen}, {Harvey}, {Evans},
  {Huard}, {Allen}, {Porras}, {Blake}, {Bourke}, {Chapman}, {Cieza}, {Koerner},
  {Lai}, {Mundy}, {Myers}, {Padgett}, {Rebull}, {Sargent}, {Spiesman},
  {Stapelfeldt}, {van Dishoeck}, {Wahhaj}, \& {Young}}]{Jorgensen_06}
{J{\o}rgensen}, J.~K., {Harvey}, P.~M., {Evans}, II, N.~J., {et~al.} 2006,
  \apj, 645, 1246

\bibitem[{{Lada}(1987)}]{Lada_87}
{Lada}, C.~J. 1987, in IAU Symposium, Vol. 115, Star Forming Regions, ed.
  M.~{Peimbert} \& J.~{Jugaku}, 1--17

\bibitem[{{Lada} {et~al.}(2006){Lada}, {Muench}, {Luhman}, {Allen}, {Hartmann},
  {Megeath}, {Myers}, {Fazio}, {Wood}, {Muzerolle}, {Rieke}, {Siegler}, \&
  {Young}}]{Lada_06}
{Lada}, C.~J., {Muench}, A.~A., {Luhman}, K.~L., {et~al.} 2006, \aj, 131, 1574

\bibitem[{{Lucas} {et~al.}(2007){Lucas}, {Hoare}, {Longmore}, {Schroder},
  {Davis}, {Adamson}, {Bandyopadhyay}, {de Grijs}, {Smith}, {Gosling},
  {Mitchison}, {Gaspar}, {Coe}, {Tamura}, {Parker}, {Irwin}, {Hambly}, {Byant},
  {Collins}, {Cross}, {Evans}, {Gonzalez-Solares}, {Hodgkin}, {Lewis}, {Read},
  {Riello}, {Sutorius}, {Lawrence}, {Drew}, \& {Dye}}]{Lucas_07}
{Lucas}, P.~W., {Hoare}, M.~G., {Longmore}, A., {et~al.} 2007, ArXiv e-prints,
  712

\bibitem[{{Luhman} {et~al.}(2006){Luhman}, {Whitney}, {Meade}, {Babler},
  {Indebetouw}, {Bracker}, \& {Churchwell}}]{Luhman_06}
{Luhman}, K.~L., {Whitney}, B.~A., {Meade}, M.~R., {et~al.} 2006, \apj, 647,
  1180

\bibitem[{{Mac Low} \& {Klessen}(2004)}]{Maclow_04}
{Mac Low}, M.-M. \& {Klessen}, R.~S. 2004, Reviews of Modern Physics, 76, 125

\bibitem[{{Makovoz} \& {Marleau}(2005)}]{Marleau_05}
{Makovoz}, D. \& {Marleau}, F.~R. 2005, \pasp, 117, 1113

\bibitem[{{Marco} {et~al.}(2001){Marco}, {Bernabeu}, \&
  {Negueruela}}]{Marco_01}
{Marco}, A., {Bernabeu}, G., \& {Negueruela}, I. 2001, \aj, 121, 2075

\bibitem[{{Marco} \& {Negueruela}(2002)}]{Marco_02}
{Marco}, A. \& {Negueruela}, I. 2002, \aap, 393, 195

\bibitem[{{Marigo} {et~al.}(2008){Marigo}, {Girardi}, {Bressan}, {Groenewegen},
  {Silva}, \& {Granato}}]{Marigo_08}
{Marigo}, P., {Girardi}, L., {Bressan}, A., {et~al.} 2008, \aap, 482, 883

\bibitem[{{Massey} {et~al.}(1995){Massey}, {Johnson}, \&
  {Degioia-Eastwood}}]{Massey_95b}
{Massey}, P., {Johnson}, K.~E., \& {Degioia-Eastwood}, K. 1995, \apj, 454, 151

\bibitem[{{Mathis} {et~al.}(1983){Mathis}, {Mezger}, \& {Panagia}}]{Mathis_83}
{Mathis}, J.~S., {Mezger}, P.~G., \& {Panagia}, N. 1983, \aap, 128, 212

\bibitem[{{Megeath} {et~al.}(2004){Megeath}, {Allen}, {Gutermuth}, {Pipher},
  {Myers}, {Calvet}, {Hartmann}, {Muzerolle}, \& {Fazio}}]{Megeath_04}
{Megeath}, S.~T., {Allen}, L.~E., {Gutermuth}, R.~A., {et~al.} 2004, \apjs,
  154, 367

\bibitem[{{Moffat} \& {Vogt}(1974)}]{Moffat_74}
{Moffat}, A.~F.~J. \& {Vogt}, N. 1974, Veroeffentlichungen des Astronomischen
  Instituts der Ruhr-Universitaet Bochum, 2, 1

\bibitem[{{Padoan} \& {Nordlund}(1999)}]{Padoan_99}
{Padoan}, P. \& {Nordlund}, {\AA}. 1999, \apj, 526, 279

\bibitem[{{Reach} {et~al.}(2005){Reach}, {Megeath}, {Cohen}, {Hora}, {Carey},
  {Surace}, {Willner}, {Barmby}, {Wilson}, {Glaccum}, {Lowrance}, {Marengo}, \&
  {Fazio}}]{Reach_05}
{Reach}, W.~T., {Megeath}, S.~T., {Cohen}, M., {et~al.} 2005, \pasp, 117, 978

\bibitem[{{Robitaille} {et~al.}(2006){Robitaille}, {Whitney}, {Indebetouw},
  {Wood}, \& {Denzmore}}]{Robitaille_06}
{Robitaille}, T.~P., {Whitney}, B.~A., {Indebetouw}, R., {Wood}, K., \&
  {Denzmore}, P. 2006, \apjs, 167, 256

\bibitem[{{Rolleston} {et~al.}(2000){Rolleston}, {Smartt}, {Dufton}, \&
  {Ryans}}]{Rolleston_00}
{Rolleston}, W.~R.~J., {Smartt}, S.~J., {Dufton}, P.~L., \& {Ryans}, R.~S.~I.
  2000, \aap, 363, 537

\bibitem[{{Sharma} {et~al.}(2007){Sharma}, {Pandey}, {Ojha}, {Chen}, {Ghosh},
  {Bhatt}, {Maheswar}, \& {Sagar}}]{Sharma_07}
{Sharma}, S., {Pandey}, A.~K., {Ojha}, D.~K., {et~al.} 2007, \mnras, 380, 1141

\bibitem[{{Siess} {et~al.}(2000){Siess}, {Dufour}, \& {Forestini}}]{Siess_00}
{Siess}, L., {Dufour}, E., \& {Forestini}, M. 2000, \aap, 358, 593

\bibitem[{{Snell} {et~al.}(2002){Snell}, {Carpenter}, \& {Heyer}}]{Snell_02}
{Snell}, R.~L., {Carpenter}, J.~M., \& {Heyer}, M.~H. 2002, \apj, 578, 229

\bibitem[{{Stern} {et~al.}(2005){Stern}, {Eisenhardt}, {Gorjian}, {Kochanek},
  {Caldwell}, {Eisenstein}, {Brodwin}, {Brown}, {Cool}, {Dey}, {Green},
  {Jannuzi}, {Murray}, {Pahre}, \& {Willner}}]{Stern_05}
{Stern}, D., {Eisenhardt}, P., {Gorjian}, V., {et~al.} 2005, \apj, 631, 163

\bibitem[{{Stetson}(1987)}]{Stetson_87}
{Stetson}, P.~B. 1987, \pasp, 99, 191

\bibitem[{{Stetson}(1994)}]{Stetson_94}
{Stetson}, P.~B. 1994, \pasp, 106, 250

\bibitem[{{Tapia} {et~al.}(1991){Tapia}, {Costero}, {Echevarria}, \&
  {Roth}}]{Tapia_91}
{Tapia}, M., {Costero}, R., {Echevarria}, J., \& {Roth}, M. 1991, \mnras, 253,
  649

\bibitem[{{Vallenari} {et~al.}(1999){Vallenari}, {Richichi}, {Carraro}, \&
  {Girardi}}]{Vallenari_99}
{Vallenari}, A., {Richichi}, A., {Carraro}, G., \& {Girardi}, L. 1999, \aap,
  349, 825

\bibitem[{{Weisskopf} {et~al.}(2002){Weisskopf}, {Brinkman}, {Canizares},
  {Garmire}, {Murray}, \& {Van Speybroeck}}]{Weisskopf_02}
{Weisskopf}, M.~C., {Brinkman}, B., {Canizares}, C., {et~al.} 2002, \pasp, 114,
  1

\bibitem[{{Wilson} \& {Matteucci}(1992)}]{Wilson_92}
{Wilson}, T.~L. \& {Matteucci}, F. 1992, \aapr, 4, 1

\bibitem[{{Wouterloot} {et~al.}(1990){Wouterloot}, {Brand}, {Burton}, \&
  {Kwee}}]{Wouterloot_90}
{Wouterloot}, J.~G.~A., {Brand}, J., {Burton}, W.~B., \& {Kwee}, K.~K. 1990,
  \aap, 230, 21

\end{thebibliography}
\end{document}